\documentclass[useAMS]{mn2e}
\usepackage{graphicx}
\usepackage{longtable}
\usepackage{amssymb}

\title[AGB Stars in the Sculptor Dwarf Spheroidal Galaxy]{Asymptotic Giant
Branch Stars in the Sculptor Dwarf Spheroidal Galaxy}
\author[Menzies et al.]{John W. Menzies$^{1}$,
Michael W. Feast$^{2,1}$,
Patricia A. Whitelock$^{1,2}$
and Noriyuki \newauthor Matsunaga$^3$\\
      $^1$ South African Astronomical Observatory, P.O.Box 9, 7935
           Observatory, South Africa.\\
      $^2$ Astronomy, Cosmology and Gravitation Centre, Astronomy Department, 
           University of Cape Town, 7701 Rondebosch, South Africa.\\
      $^3$ Kiso Observatory, Institute of Astronomy, The University of Tokyo,
           10762-30, Mitake, Kiso, Nagano 397-0101, Japan.\\
 }
\begin{document}
\maketitle

\begin{abstract} 
$JHK_S$ photometry is presented for a 35 arcmin square field centred on the Sculptor dwarf spheroidal galaxy. With the aid of published kinematic data definite galaxy members are identified and the width in $J-K$ of the colour-magnitude diagram is shown to be consistent with an old population of stars with a large range in metal abundance. We identify two Asymptotic Giant Branch variables,
both carbon Miras, with periods of 189 and 554 days, respectively, and discuss
their ages, metallicities and mass loss as well as their positions in the Mira period--luminosity diagram. 

\noindent There is evidence for a general period--age relation for Local Group Miras. The mass-loss rate for the 554-day variable, MAG29, appears to be consistent with that found for Miras of comparable period in other Local Group galaxies.

\end{abstract}

\section{Introduction}
 This paper is one of a series aimed at finding and characterizing luminous
Asymptotic Giant Branch (AGB) variables within Local Group galaxies; it
follows similar work on Phoenix, Fornax  and Leo I
(Menzies et al. 2002, 2008; Whitelock et al. 2009; Menzies et al. 2010).

Sculptor was the first dwarf spheroidal galaxy to be discovered (Shapley 1938).
Variable stars were found in it by Baade \& Hubble (1939) and Thackeray (1950)
confirmed that these were mainly RR Lyrae variables.
Establishing the nature of these variables was important for Baade's general
concept of populations I and II (see letter from Baade to Thackeray 1950 Mar 27
in Feast 2000).
Further work on the variables
was carried out by van Agt (1978) and Kaluzny et al. (1995). Most of the variables
are RR Lyrae stars but van Agt also discovered some red variables one of which
we show below to be a Mira variable. 

Carbon stars were discovered in the system
by Azzopardi, Lequeux \& Westerlund (1985, 1986), while Frogel et al. (1982) reported another three possible examples. In more recent times the
system has been extensively studied for stellar chemical abundances (e.g. 
Coleman et al. 2005, Battaglia et al. 2008, Kirby et al. 2009) 
kinematics (e.g. Westfall et al. 2006; Walker et al. 2009) and star formation rate (Hurley-Keller 2000). 

Sculptor is believed to contain
a predominantly old population but with a small population of age $\sim 2$ Gyr and less
(Revaz et al.  2009, especially their fig. 13). Work by various authors has shown that there is
a large range in [Fe/H]. Kirby et al. (2009)  found a mean value of $-1.58$  and an asymmetric tail
extending out to $\sim -3.0$. There is evidence of a metallicity gradient
in the galaxy with the more metal strong stars preferentially concentrated to the centre
(see Tolstoy et al. 2004; Westfall et al. 2006; Walker et al. 2009; Kirby et al. 2009).

\section{Observations}
 Observations were made with the SIRIUS camera on the Japanese-South African
Infrared Survey Facility (IRSF) at SAAO, Sutherland. The camera produces
simultaneous $J$, $H$ and $K_S$ images covering an approximately $7.2 \times
7.2$ arcmin square field (after dithering) with a scale of 0.45 arcsec/pixel.
Images were obtained on a $5\times 5$ grid centred on the nominal centre of the galaxy, 01:00:06,
--33:44:13 (2000.0) (van Agt 1978). With overlaps between adjacent
fields, the area observed was approximately 35 arcmin squared. An additional field was included later to allow one of the carbon stars (ALW1) to be observed.

Images were obtained at 25 epochs spread over 4 years in the central 9 fields, and between 16 and 20 epochs in the remaining
fields. In each field, 10 dithered frames were combined after
flat-fielding and dark and sky subtraction. Exposure times were either
30 or 20\,s for each frame, depending on the seeing and on the brightness of
the sky in the $K_S$ band. Photometry was performed using DoPHOT (Schechter
et al. 1993) in `fixed-position' mode, using the best-seeing $H$-band image
in each field as templates. Aladin was used to correct the coordinate system
on each template and RA and Dec were determined for each measured star. This
allowed a cross-correlation to be made with the 2MASS catalogue (Cutri et
al. 2003), and photometric zero points were determined by comparison of our
photometry with that of 2MASS. In each field, stars in common with the 2MASS
catalogue with photometric quality A in each colour were identified and the
IRSF zero point was adjusted to match that of 2MASS. The mean number of common
stars per field was 12 and the 
mean standard deviation over all fields of the differences
between IRSF and 2MASS are 0.04 mag in $J$ and $H$ and 0.06 mag in
$K_S$. No account was taken of possible colour transformations, such as in
Kato et al. (2007). Those transformations were derived using highly reddened
objects to define the red end and it is not obvious that the same
transformations will apply to carbon stars.

\section{Colour-Magnitude and Colour-Colour Diagrams}
\begin{figure} 
\includegraphics[width=8.5cm]{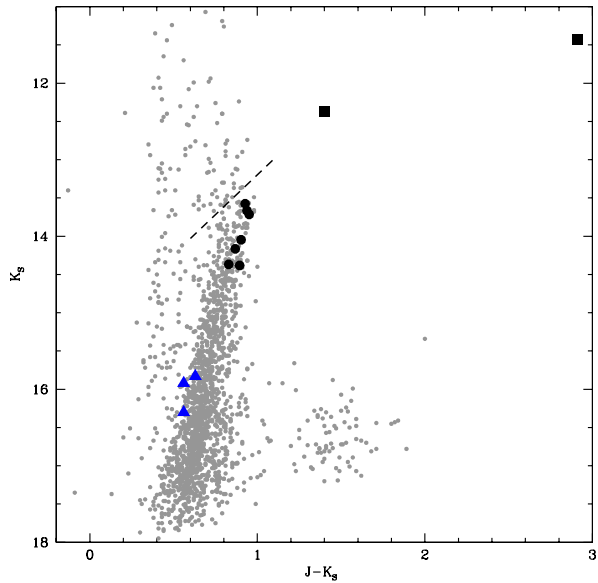}
\caption{Colour-magnitude diagram for all stars measured in our Sculptor field; small grey dots are field and 
red
giant branch stars; squares are variables. Black points are carbon stars. Two carbon stars not observed by us are plotted at their 2MASS positions. Three stars with [Fe/H]$<-3.4$ are shown as blue triangles. The dashed line shows the approximate expected position of the tip of the red giant branch assuming a distance modulus of 19.46 mag (see text).}

\label{fig_cm} 
\end{figure}

\begin{figure}
\includegraphics[width=8.5cm]{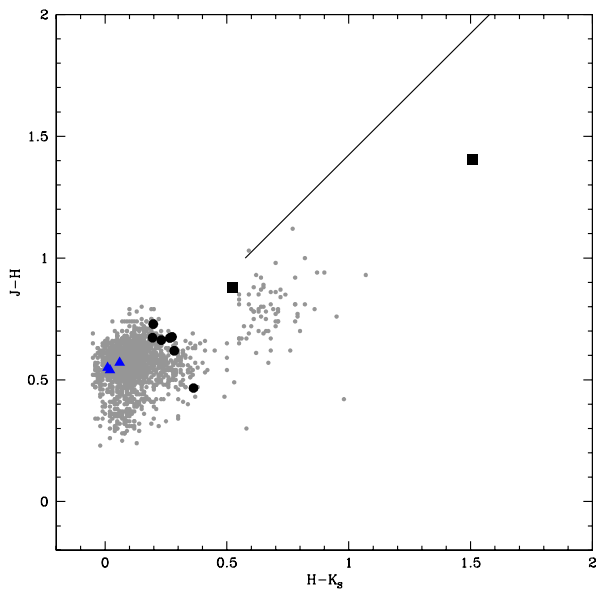}
\caption{Two-colour diagram for Sculptor; symbols as in Fig.~\ref{fig_cm}. The   solid line represents the locus of
Galactic carbon Miras (see Whitelock et al. 2009).}
\label{fig_jhhk}
\end{figure}

Fig.~\ref{fig_cm} shows the $K_S-(J-K_S)$ diagram and
Fig.~\ref{fig_jhhk} the $(J-H)-(H-K_S)$ diagram for all the stars discussed in this paper.
%two variables, and for constant stars selected on the basis of their standard deviations. 
Mean magnitudes from all epochs are used in the plots.

Following the interpretation of the similar diagram for the Fornax dwarf
spheroidal, the main features of Fig.~\ref{fig_cm} are: a well defined RGB with a tip near $K = 13.5$ mag; bluer
stars of all magnitudes which are likely foreground stars; a few stars above the TRGB which
are likely foreground stars, though some might be AGB members; two bright red variables; unresolved background
galaxies with $J-K \sim 1.5$ mag extending from $K\sim 16.0$  mag to fainter magnitudes.

Bellazzini (2008) gives a relation between M$_K$ and $(J-K)$ for the tip of the red giant branch (TRGB) which is shown as a line in
Fig.~\ref{fig_cm}, assuming a distance modulus of 19.46 mag and adjusted for consistency with an LMC distance modulus of 18.39 mag . It is interesting to note that, apart from the variables, all the C stars lie below the TRGB. Three stars with [Fe/H]$<-3.4$ are shown in the figure, and their details are listed in Table~\ref{tab_C} for convenience.

\begin{figure}
\includegraphics[width=8.5cm]{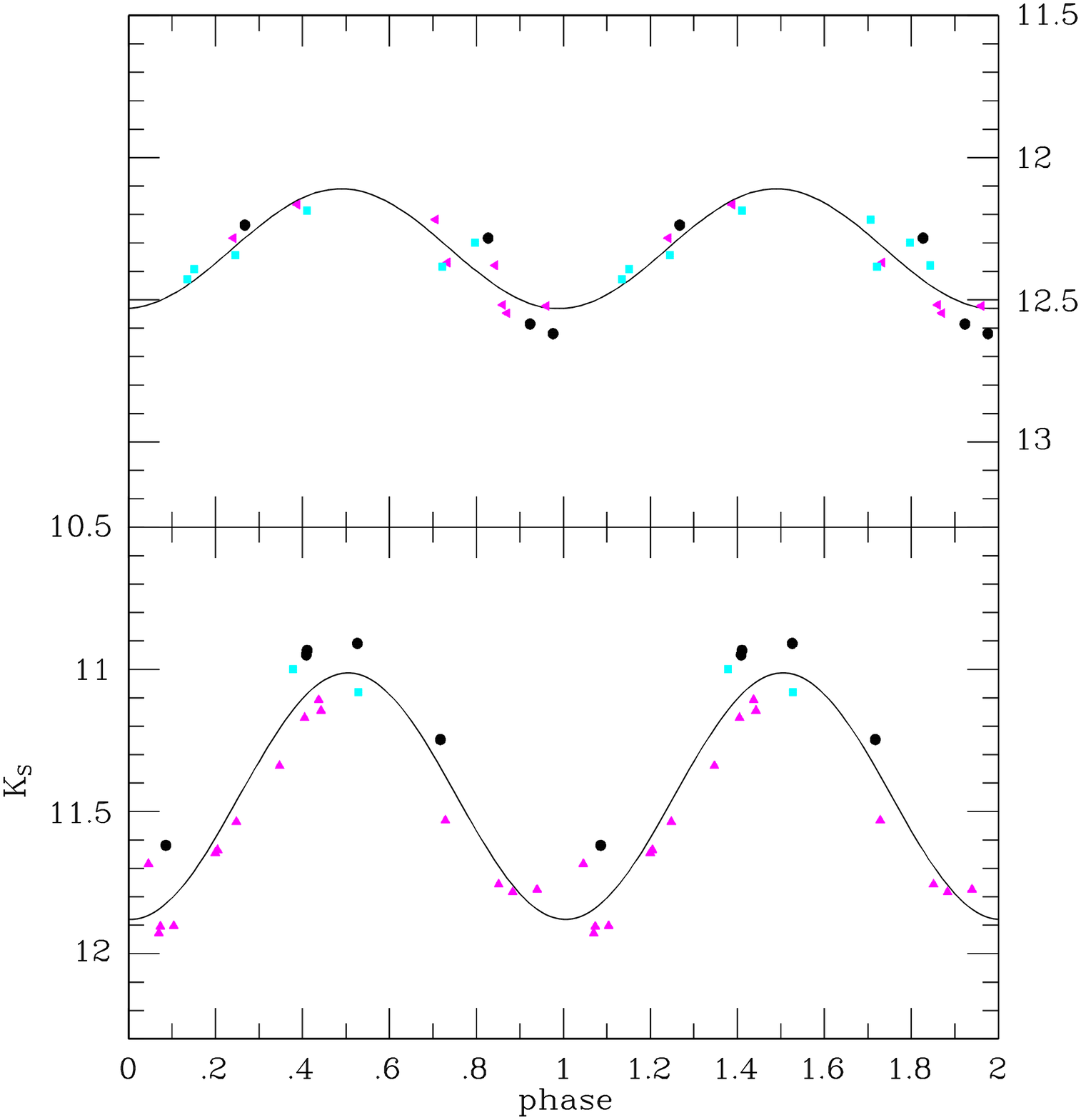}
 \caption{$K_S$ light curves of (bottom panel) MAG29 and (top panel) V544
phased with periods 554 days and 189 days, respectively, and arbitrary epoch
of JD 2450000. For MAG29 filled circles (black) show first pulsation cycle,
triangles (magenta) show second cycle, squares (cyan) show third cycle. For V544, these same
symbols show the first, second, third {\em pairs} of cycles, respectively.}
 \label{fig_var}
\end{figure}

We have found only two periodic red variables in Sculptor; one is our catalogue number 12002, found to be variable by van Agt (1978) and catalogued as V544. The other is newly discovered (our 3001), and was found to be a C star, designated MAG29, in a survey of 2MASS red stars by Mauron et al. (2004). The light curves are shown in Fig.~\ref{fig_var}. 
Following Whitelock et al. (2006) we classify them both as Mira variables, because they are periodic and have amplitudes $\Delta K > 0.4$ mag;
MAG29 has a period of 554 days, and V544 a period of 189 days. There is some evidence for long term variations in the MAG29 lightcurve, but with only about two cycles
covered by our observations it is difficult to comment constructively on this point; more data are required.

%\section{Carbon stars and Variables}
\onecolumn

\begin{table}
\caption[]{Red variables, Carbon stars  and very metal-poor stars in Sculptor.}
\begin{tabular}{cccccccllll}
\label{tab_C} \\
\hline
\multicolumn{1}{c}{$\alpha$} & $\delta$ & Scl\# & $J$ & $H$ & $K_S$ & $J-K_S$ &  other name&\multicolumn{2}{c}{Membership}&note\\
\multicolumn{2}{c}{(equinox 2000.0)}& & \multicolumn{4}{c}{------------------(mag)------------------}& &&Ref&\\
\hline
\multicolumn{10}{l}{\it Variables} \\
00:59:53.8 & --33:38:31 &  3001 &  14.35 &  12.94 &  11.44 &  2.91 & MAG29 &&&1\\
00:59:58.9 & --33:28:35 & 12002 &  13.78 &  12.90 &  12.38 &  1.40 & V544, ALW3 &&&2,3\\
\multicolumn{8}{l}{\it Carbon stars} \\
00:58:36.1 & --33:40:26 & 26010 &  14.67 &  13.99 &  13.72 &  0.95 & ALW1 &Y&C&3\\
01:00:10.4 & --33:51:00 &  7010 &  15.28 &  14.61 &  14.38 &  0.89 & ALW2 &Y&C&3\\
01:00:10.7 & --33:40:58 &  1023 &  14.95 &  14.33 &  14.05 &  0.90 & ALW4, FBMC12 &Y&W&3,4\\
01:00:20.7 & --33:45:18 &  1003 &  14.60 &  13.93 &  13.66 &  0.94 & ALW5, FBMC4 &Y&S&3,4\\
01:00:27.4 & --33:53:04 &  8008 &  15.03 &  14.36 &  14.17 &  0.87 & ALW6 &&&3\\
00:58:27.1 & --34:01:19 &     & 14.50 & 13.77 & 13.57 &  0.93  &   ALW7 & Y & C&3,5 \\
00:58:30.7 & --33:28:40 &     &  15.20 & 14.73 & 14.37 &  0.83  &   ALW8 & Y & W &3,5 \\
\multicolumn{8}{l}{\it Metal-poor stars} \\
01:00:47.8 & --33:41:03 &  2030 & 16.46 & 15.89 & 15.83 &  0.63 &S1020549  & Y &S K & 6 \\
01:00:05.0 & --34:01:17 & 20028 & 16.48 & 15.94 & 15.92 &  0.56 &Scl07-49  & Y & Wa & 6 \\
01:00:01.1 & --33:59:21 & 20034 & 16.86 & 16.31 & 16.30 &  0.56 &Scl07-50  & Y & We & 6 \\
\hline

\multicolumn{10}{l}{{\bf Notes:}}\\
\multicolumn{10}{l}{(1) MAG29 is 2MASS00595367--3338308. It is a C star according to Mauron et al. (2004)} \\
\multicolumn{10}{l}{ MAG29 has a peak-to-peak K amplitude of 0.87 mag.} \\
\multicolumn{10}{l}{(2) V544 from van Agt (1978). Peak-to-peak K amplitude 0.42 mag.}\\
\multicolumn{10}{l}{(3) ALW numbers from Azzopardi et al. (1986). Note ALW2 and 3 are reversed in Azzopardi et
al. (1985).} \\
\multicolumn{10}{l}{(4) FBMC numbers from Frogel et al. (1982).} \\
\multicolumn{10}{l}{(5) 2MASS data for ALW7 (J00582732--3401186) and ALW 8
(J00583082--3328401) outside our field.}\\
\multicolumn{10}{l}{(6) [Fe/H]: --3.81 (Scl2030) Frebel et al. (2010); --3.48 (Scl20034),
--3.96 (Scl20028) (Tafelmeyer et al. 2010).}\\ 
\multicolumn{10}{l}{Member (Y) according to: Coleman et al. (2005): C, Schweitzer et al. (1995): S, Walker et al. (2009): Wa,} \\
\multicolumn{10}{l}{Westfall et al. (2006): We.} \\

\end{tabular}
\end{table}

%\begin{center}
%\onecolumn
\begin{table}
\caption[]{$JHK_S$ for stars in Sculptor.}
\begin{tabular}{cccccccl}
\label{tab_memb} \\
\hline
$\alpha$ & $\delta$ & Scl\# & $J$ & $H$ & $K_S$ & $J-K_S$ & Memb \\
\multicolumn{2}{c}{(equinox 2000.0)}& & \multicolumn{4}{c}{-------------- (mag) -----------------}&  \\
\hline

\multicolumn{8}{l}{\it Members}\\
00:58:15.7 & -33:40:27 & 26060 & 17.32 & 16.77 & 16.77 &  0.56 &  Wa   \\ 
00:58:19.5 & -33:43:00 & 26038 & 16.66 & 16.05 & 16.03 &  0.63 &  Wa We   \\ 
00:58:29.7 & -33:38:22 & 26064 & 17.29 & 16.68 & 16.64 &  0.66 &  Wa   \\ 
00:58:33.7 & -33:40:42 & 26043 & 17.25 & 16.63 & 16.62 &  0.63 &  Wa   \\ 
00:58:33.7 & -33:43:18 & 26013 & 15.15 & 14.48 & 14.34 &  0.81 &  C We   \\ 
00:58:36.1 & -33:40:26 & 26010 & 14.67 & 13.99 & 13.72 &  0.95 &  C   \\ 
00:58:38.0 & -33:43:03 & 26055 & 17.41 & 16.87 & 16.74 &  0.66 &  Wa   \\ 
00:58:39.2 & -33:43:14 & 26054 & 17.35 & 16.83 & 16.74 &  0.62 &  Wa   \\ 
00:58:40.8 & -33:39:44 & 26044 & 17.17 & 16.58 & 16.54 &  0.63 &  Wa   \\ 
00:58:43.1 & -33:43:34 & 26036 & 16.79 & 16.24 & 16.14 &  0.65 &  Wa   \\ 
00:58:43.4 & -33:35:59 & 15035 & 16.67 & 16.11 & 16.03 &  0.63 &  Wa C   \\ 
00:58:44.0 & -33:41:45 & 16086 & 17.09 & 16.55 & 16.43 &  0.66 &  Wa   \\ 
00:58:44.7 & -33:41:56 & 16006 & 14.66 & 14.00 & 13.88 &  0.78 &  We   \\ 
00:58:44.7 & -33:30:49 & 14018 & 16.31 & 15.74 & 15.61 &  0.70 &  Wa We   \\ 
00:58:45.5 & -33:38:02 & 15026 & 16.73 & 16.19 & 16.16 &  0.57 &  Wa   \\ 
00:58:45.6 & -33:35:10 & 15059 & 17.61 & 17.16 & 17.09 &  0.52 &  Wa   \\ 

\hline
\multicolumn{8}{l}{Notes: }\\
\multicolumn{8}{l}{Membership references: Battaglia et al. (2008): B, Coleman et al. (2005): C,}\\
\multicolumn{8}{l}{Kirby et al. (2009): K, Schweitzer et al. (1995): S, Walker et al. (2009): Wa,}  \\
\multicolumn{8}{l}{Westfall et al. (2006): We.}  \\
\multicolumn{8}{l}{This is a partial table to indicate the format and contents. The full table} \\
\multicolumn{8}{l}{including definite members and non-members, as well as stars with unknown status} \\
\multicolumn{8}{l}{is available as an attachment in the online version.}\\
\end{tabular}
\end{table}
%\end{center}
%\twocolumn

\twocolumn

Of the eight carbon stars identified by Azzopardi et al. (1986), six fall in the area of our survey. In a study of late-type stars in Sculptor, 
Frogel et al. (1982) find two definite C stars (in common with Azzopardi et al.) and three possible ones, which however were not confirmed by 
Richer \& Westerlund (1983).
Our data for the variables and C stars are given in Table~\ref{tab_C}.

\begin{figure}
\includegraphics[width=8.5cm]{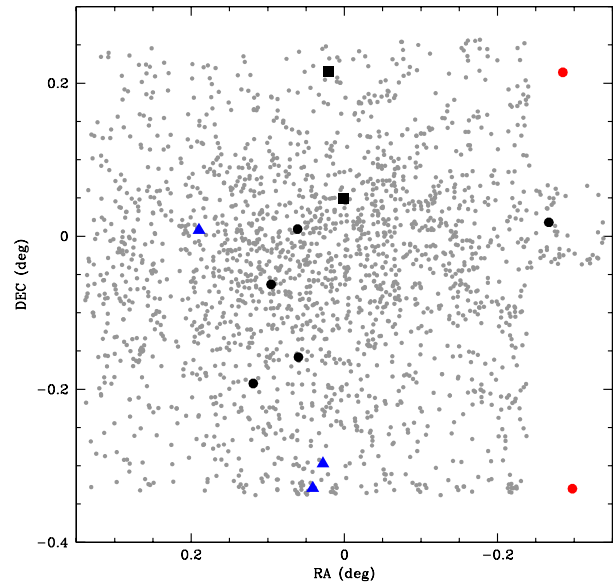}
\caption{All stars observed in the Sculptor field (grey dots). Variables are filled squares (the northern one is V544, the one near the centre of the field is MAG29). Carbon stars are large black symbols except for two carbon stars that fall outside our observed field that are indicated by red filled circles on the right side of the figure. Three stars
with very low metallicities are shown as blue triangles.} 
\label{fig_field}
\end{figure}

The distribution on the sky of the stars included in the colour-magnitude diagram is illustrated in
Fig.~\ref{fig_field}. 

%\begin{center}
\section{Membership and Metallicity}
%NOTE. IN DISCUSSING METALLICITIES FROM CaT WE NEED TO TAKE INTO ACCOUNT
%WHERE NECESSARY ANY RECALLIBRATION AT THE LOW METALLICITY END
%(SEE FOR INSTANCE TAFELMEYER FIG 6)
% Interpretation of the colour-magnitude and colour-colour
%diagrams for Sculptor are greatly helped by the studies of
%proper motion, radial velocity and chemical abundance that have been
%carried out. The figures in this section combine our photometry
%with these studies. Care is needed in discussing these figures since
%the selection of stars plotted tend to be limited either by the
%area studied or the way the stars measured were selected.
Babusiaux et al. (2005) have drawn attention to the considerable breadth in the infrared of Sculptor's  giant branch. They deduce a range of 0.75 dex in [Fe/H] on the basis of fits of Padova theoretical isochrones for a fixed age to the $K_S$, $V-K_S$ colour-magnitude diagram. Since then, a number of kinematic and membership studies have been conducted, some of which also determine metallicities for individual stars. We use these new results to investigate the star formation history of Sculptor further. 

A proper-motion study covering 1177 stars within a 7.25 arcmin radius of the Sculptor centre was conducted by Schweitzer et al. (1995). Our limiting magnitude is significantly brighter than theirs, but we have observed 407 star in common with their presumably unbiased selection, including 381 member stars.

Coleman et al. (2005) conducted a wide-area (10 square degrees centred on the galaxy) photometric and spectroscopic survey of Sculptor with a view to mapping the outer structure of the galaxy. The $V,V-I$ colour-magnitude diagram was used to select likely members. Velocities were determined from the spectra and the equivalent width of two Ca II triplet lines used to determine [Fe/H] on the Zinn \& West (1984) metallicity scale. Data were only published for members, of which 84 are in common with our photometry.

Another wide-area investigation was conducted by Westfall et al. (2006), this time employing Washington $M, T2 + DDO51$ photometry to isolate likely members. Their final catalogue contains 179 stars (157 velocity members) of which 113 are in common with our catalogue. No metallicity information was provided.

Sculptor was included in a velocity survey of four Local Group dwarf spheroidals by Walker et al. (2009). The area covered extends well beyond our survey fields, and a total of 1541 stars with velocities and line strength information are catalogued. These were selected on the basis of their positions within a box drawn on a $V,V-I$ colour-magnitude diagram. The bright limit appears to lie well below the tip of the red giant branch. Our survey has 762 stars in common with Walker et al., including 729 members. 

In a large-scale survey of abundances in Sculptor, Kirby et al. (2009) catalogue [Fe/H] values based on Ca II triplet measurements for 388 stars which were selected from the Westfall et al. (2006) paper. Only radial-velocity members are listed, but the underlying sample is subject to the same selection criteria as in the Westfall et al. study. Kirby et al. show that their abundances are in good agreement with those derived from high-resolution spectra by Battaglia et al. (2008).

Finally, in a paper aimed at comparing Ca II triplet and high-resolution spectroscopy of Sculptor stars, Battaglia et al. (2008) catalogue 93 stars, all of which are in common with our survey. They show that Ca II triplet abundances calibrated with respect to globular cluster [Fe/H] data give a reliable measure of metal abundance for members of Sculptor. 

Taking all of these surveys together, we find 952 definite members and 48 non-members in our sample, with a further 796 stars of unknown status. Our final $JHK_S$ 
data, with membership status, are listed in Table~\ref{tab_memb}. The coordinates are as determined by us, and the column labelled ``Scl\#" refers to our internal numbers for the stars measured.

\begin{figure}
\includegraphics[width=8.5cm]{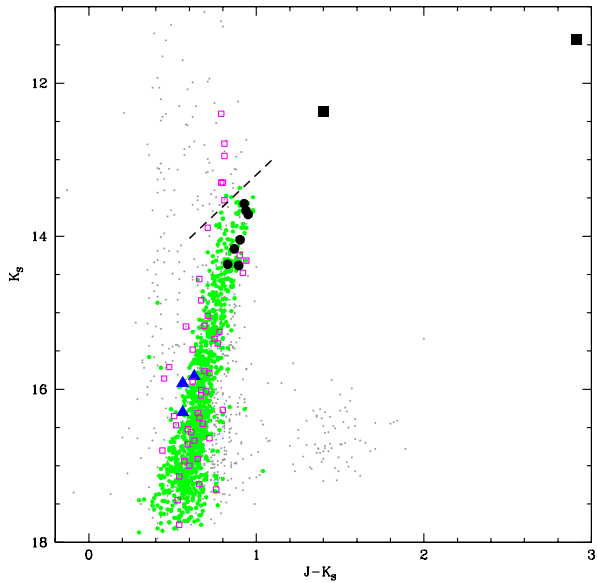}
\caption{Colour-magnitude diagram for Sculptor with symbols as in Fig.~\ref{fig_cm}.
The line shows the expected position of the tip of the red giant branch assuming a distance modulus of 19.46 mag (see text). Definite proper motion or radial velocity members are overplotted as green dots, while non-members are shown as magenta open
squares.}

\label{memb_cm}
\end{figure}

The $K_S, J-K_S$ colour-magnitude diagram is shown in Fig.~\ref{memb_cm}, with the definite members marked in green, and definite non-members in magenta. Given the biases in the catalogues referred to above, too much should not be read into the distribution of stars with $K_S$ magnitude, especially at the bright end where AGB stars, if any, have probably been discriminated against. The sloping line indicates where the tip of the red giant branch is expected to be (Bellazzini 2008) for a range of $J-K_S$ magnitudes as proxies for metal abundance. 

It is noteworthy that some of the `definite' members lie some distance from the giant branch in Fig.~\ref{memb_cm} on the blue side. There are three outliers more than 0.1 mag bluer than the blue edge of the giant branch. Three stars from the Schweitzer et al (1995) catalogue, our numbers 1017, 2034 and 9033, have probabilities of membership of 0.61, 0.90 and 0.78, respectively. They all lie in a part of
the $J-H, H-K$ diagram that suggests they are most likely field stars, but it would be worth checking on this. On the red side, star 7082 is about 0.2 mag redder than the red edge of the giant branch, but is a member according to Walker et al. (2009).

\begin{figure}
\includegraphics[trim=0mm 0mm 0mm 82mm, clip, width=8.5cm]{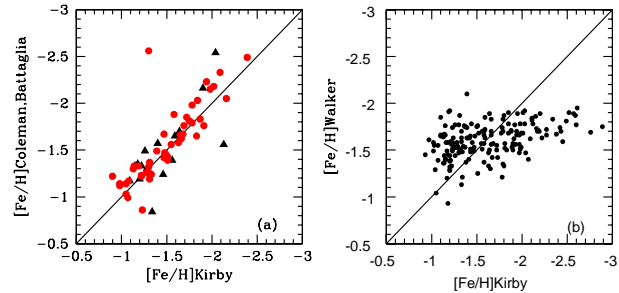}
\caption{(a) Comparison of [Fe/H] for Battaglia et al. (2008)(dots) and Coleman et al. (2005)(triangles) with those of Kirby et al. (2009) for stars in common with our survey. (b) Similar comparison for Walker et al. (2009)([Fe/H] from Mg indices) and Kirby et al. (2009) data.}

\label{fig_spec}
\end{figure}

We have already alluded to the good agreement between the Kirby et al. (2009) and Battaglia et al. (2008) metallicities. The same applies to the Coleman et al. (2005) results, 
once they have been converted to the Carretta \& Gratton (1997) scale. This is
illustrated in Fig.~\ref{fig_spec}(a). Walker et al. (2009) measured equivalent widths for several Fe and Mg absorption lines in their spectra. They calibrated these with respect to globular cluster abundances to give [Fe/H] for the Sculptor stars. They cautioned against the use of these as absolute abundances since in two other  Local Group galaxies, abundances derived from Fe and Mg lines in this way gave rise to significantly narrower ranges than did those derived from the Ca II triplet. We find the same result for stars we have in common with both Kirby et al. (2009) and Walker et al. (2009) as shown in Fig.~\ref{fig_spec}(b). 

We have determined the mean $J-K_S$ as a function of $K_S$ for the giant branch members and have divided them into two groups, redder or bluer than the mean. Using the metallicities from Coleman et al. (2005), Kirby et al. (2009) and Battaglia et al. (2008) we find a mean
[Fe/H] of --1.48 for 193 giants on the red side, and [Fe/H] of --1.81 for 184 giants on the blue side. The two distributions are relatively broad and overlap, but the large range of abundances across the giant branch is clear.

\begin{figure}
\includegraphics[width=8.5cm]{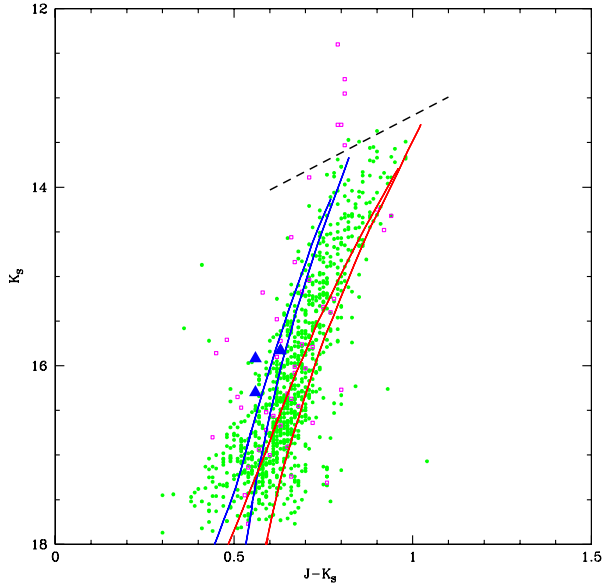}
\caption{Colour-magnitude diagram for Sculptor members. Symbols as in Fig.~\ref{memb_cm}.
Isochrones for 12.7 Gyr populations with [Fe/H]=--1.76 (blue lines) and --1.4 (red lines) are overplotted.}

\label{oldpop}
\end{figure}

In Fig.~\ref{oldpop} we show the colour-magnitude diagram for the members and C stars (MAG29 is off the red edge). Superposed on the plot are isochrones (Marigo et al. 2008) for populations of age 12.7 Gyr and metal abundances, [Fe/H] = --1.76 (blue) and --1.4 (red), to illustrate that the width of the giant branch can be explained this way. Each isochrone has two parts, the rightmost one representing stars on the RGB,  and the other the stars on the AGB.

\begin{figure}
\includegraphics[width=8.5cm]{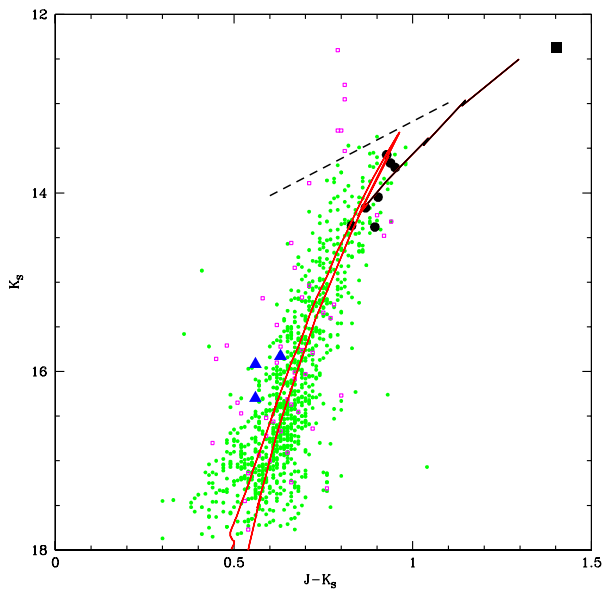}
\caption{Colour-magnitude diagram for Sculptor members. Symbols as in Fig.~\ref{memb_cm}.
 An isochrone (Marigo et al. 2008) for a population of age 2 Gyr and [Fe/H] $\sim -1.1$
is overplotted as red lines. The region on the isochrone where C/O $>$ 1 is shown in black.}

\label{young}
\end{figure}

The presence of two very red Miras, both carbon stars,  indicates that there must be an intermediate age population in Sculptor as well as the old population (see section 6). It is puzzling that the other carbon stars lie below the expected TRGB in the colour-magnitude diagram. However, Revaz et al. (2009) have found evidence for a population of age $\sim$2 Gyr. In 
Fig.~\ref{young} we show an isochrone for a 2 Gyr population and [Fe/H] $\sim -1.1$ superposed on the colour-magnitude diagram of members. The section of the isochrone marked in black shows where the C/O ratio exceed 1.0, which suggests that carbon stars can be produced in the region of the diagram where they are observed if such a young relatively metal-rich population exists in Sculptor.

%\section{Asymptotic Giant Branch}

%\section{Giant Branch}
%  Coleman at al. (2005) determined membership of Sculptor based on radial velocities and proper motions. They found metallicities for these stars from the Ca II triplet. The colour-magnitude diagram for stars in common with the Coleman et al. study is shown in Fig.~\ref{fig_cole} where different symbols are used for
%stars with [Fe/H] greater and less than -1.85.

%
%\begin{figure} 
%\includegraphics[width=8.5cm]{cole.ps}
%\caption{Colour-magnitude diagram for stars in common with Coleman at al. red dots ares tars with [Fe/H]$>$-1.85 and blue dots for [Fe/H]$<$-1.85.
%}
%\label{fig_cole} 
%\end{figure}

\section{Miras and the distance to Sculptor}
  
Table~\ref{tab_dist} lists various estimates of the distance to Sculptor. The distances
of the two carbon-rich Miras are based on the $K$ and/or $M_{bol}$ period--luminosity (PL)
relations:\\
\begin{equation}M_K  = -3.52[log P -2.38] -7.24    \end{equation}
from Whitelock, Feast and van Leeuwen (2008), and \\
\begin{equation} M_{bol} = -3.31[ log P -2.5] -4.271 \end{equation}
from Whitelock et al. (2009).

Both these relations were derived from carbon-rich Miras in the LMC and assume
a modulus of 18.39 mag obtained from a classical Cepheid distance scale based
on parallaxes of Galactic Cepheids and with a metallicity correction
applied to the LMC Cepheid results (van Leeuwen et al. 2007).
In the case of the 554-day Mira only the $M_{bol}$ result is listed.
In view of the redness of the star ($J-K = 2.89$) mag we anticipate significant 
circumstellar absorption at $K$ and indeed the apparent distance modulus 
given using  Eq. (2) is 0.4 mag greater than the $M_{bol}$ modulus. In the case of 
the 189-day variable, Eq. (1) gives a modulus of 19.20 mag and Eq. (2)  a modulus of 19.18 mag. 
Data for Sculptor RR Lyraes in $K$ were taken from Pietrzy\'nski et al. (2008). The
$V$ data for the Sculptor RR Lyraes 
are from 
Kaluzny et al. (1995). A reddening of $E(B-V) = 0.018$ mag from 
Schlegel et al. (1998) is adopted (see Pietrzy\'nski et al. 2008).
%[Presumably no reddening has been put into Miras yet?]. 
The derivation of the PL($K$) and M($V$)--[Fe/H] relations used is discussed by 
Feast (2010). %(Moscow vol, arXiv:0912.4159).   
The table also gives the LMC moduli obtained using the same relations.
A recent discussion of the TRGB distance scale (Feast 2011) gives a 
distance modulus of 19.43 mag for the LMC and 19.50 mag for Sculptor (using data from 
Sakai et al. (2000) and Rizzi et al. (2007)). Since we are 
here concerned primarily with the relative results of different indicators we 
scale the Sculptor modulus in the table to an LMC value of 18.39 mag.

\begin{table}
\caption[]{Distance modulus determinations for Sculptor.}
\begin{tabular}{lcc}
\label{tab_dist} \\
\hline
 Method &  Sculptor  &   LMC     \\
\hline
554 d C Mira ($M_{bol}$)     &   19.57  &   (18.39)    \\
189 d C Mira ($M_{bol}/M_K$) &  19.19  &   (18.39)  \\
  \\
RR Lyraes ($K$)       &      19.37  &    18.37  \\
RR Lyraes ($V$)       &      19.46  &    18.38  \\
  \\
TRGB($VI$)            &      19.46  &    (18.39)  \\

\hline
\end{tabular}
\end{table}

The mean of the last three values is 19.43 mag and the mean of
the two Miras is 19.38 mag, so the agreement is satisfactory.
There is some concern that the 189-day variable gives a modulus
0.24 mag smaller than the mean of the non-Mira distance indicators.
However, the scatter in the LMC PL relations is 0.15 mag for $K$ and 0.12 mag
in $M_{bol}$ (see Whitelock et al. 2008, 2009), so the modulus difference is 
1.6 to 2.0 $\sigma$ and may not be significant. 
One might suspect a difference could result from a metallicity 
effect on the zero-points, especially as the 189-day variable is likely 
to be very old (see below) and metal poor. The LMC C-Mira relations are 
poorly defined at the 
short period end. However, short period Miras in Leo~I and Fornax fit them well 
(Menzies et al. 2010,  Whitelock et al. 2009). 

\section{Ages of carbon-rich Miras}

\begin{table}
\caption[]{Metal abundances in Local Group galaxies.}
\begin{tabular}{lll}
\label{tab_abund} \\
\hline
     Galaxy    &    [Fe/H]        &      Mira Periods (days)   \\
\hline
  Phoenix      &   --1.37/--1.8 (old pop)          & 425   \\
  Leo I        &   --1.4                          & 158, 180, 191, 252,\\
               &                                 & 283, 336, 523   \\
  Fornax       &   --0.9 (--2.0 to --0.4)           & 215, 258, 267, 280,\\
               &                                 & 350, 400, 470   \\
  Sculptor     &   --1.6 (14 Gyr) to --1.2 (4 Gyr)  & 189, 554    \\

\hline
\end{tabular}
\end{table}
%[I need to look further at the best metallicities to quote]

%Our work on Sculptor and three other dwarf spheroidals of the Local Group allows us to
%put some constraints on the relation of C-Mira ages to their periods and on the  possible
%dependence of PL slope and zero-point on metallicity.
Mira variables have considerable potential both as Galactic and extragalactic distance
indicators and as indicators of the age of stellar systems. Our work on dwarf spheroidals
of the Local Group allows us to test Mira PL relations for possible
metallicity dependence and to check whether our results are consistent with an period--age
relation.

For oxygen-rich Miras the evidence (summarized by Feast 2009) indicates that  
periods increase with decreasing age and increasing initial mass, from $\sim150$ days at 
an age of $\sim12$ Gyr to 450 days at $\sim3$ Gyr and OH/IR Miras with periods 
$\sim1000$ days are even younger. A broadly similar trend of period with age is found 
amongst the longer period carbon-rich Miras.  Thus Galactic kinematics (Feast et al. 2006) 
indicates an age of $\sim1.8$ Gyr for carbon-rich Miras of period $\sim520$ days with some indication 
of an increasing period with decreasing age. 
There is a Mira with a period of 450 days in
NGC 1783 for which  Milone et al. (2009) give an age of $\sim 1.5$ Gyr,
%There is a 526-day C-Mira in the SMC cluster 
%NGC419 for which Girardi et al. (2009) give an age of $\sim1.4$ Gyr. 
%Ages of $\sim1.5$ Gyr and 2.0Gyr have been given for the LMC clusters NGC 1783 and 
%NGC1978 which contain Miras of periods 450 and 491 days respectively 
%(Milone et al. 2009). 
and if a carbon-rich Mira of period 680 days is a member 
of an LMC cluster (van Loon et al. 2003) its age is $\sim 1.0$ Gyr. 
It is likely that there will be a range in periods at a given age. How large
this is is not precisely known. The cluster NGC 1978 (age $\sim 2$ Gyr (Milone
et al. 2009)) contains two carbon-rich Miras
for which Kamath et al. (2010) give periods of 376 and 458 days, while Nishida et al. 
give 491 days for the second star. In NGC~419 there are also two Miras (periods 488 and 738 days
(Kamath et al 2010) for the first of which Nishida et al. give a period of 526 days).
The mean age of this cluster is $\sim 1.5$ Gyr though there appears to have been an
extended period of star formation in it (Rubele, Kerber \& Girardi 2010), which
seems a common phenomenon in intermediate age clusters in the Magellanic Clouds
(Milone et al. 2009). A more detailed discussion of the
evolution of AGB variables will be given elsewhere, but it will be clear that Mira
variables with periods $\sim 200$ days and lying on the PL relations with $M_{bol} \sim -3.6$ mag
are far below the RGB tip of clusters such as NGC1978 and must presumably have
evolved from lower mass (older) stars.

Table~\ref{tab_abund} lists the periods of the Miras in the four Local Group galaxies we have so
far studied. All these Miras are known to be, or are very likely to be carbon rich. We can now ask
how consistent are these Local Group results with the period--age correlation just described.

%These results can be used, together with the results listed in Table~\ref{tab_abund} to discuss the 
%implications for the stellar populations of these galaxies. Miras are rare objects and 
%in view of the small number involved, it should be remembered that whilst the presence of
%a single Mira may be taken as evidence of a significant progenitor population, absence
%does not necessarily imply absence of that population.
Consider first the long period Miras ($\sim400 - 550$ days). These imply  a significant
population of a few Gyr age in all the systems. This is consistent with other evidence
for Phoenix, Leo~I and Fornax. 

In Phoenix there is evidence of significant star formation
in the age range 1 to 5 Gyr (see, e.g. Young et al. 2007, fig.~6)
and a lower rate between 5 and 10 Gyr. In Leo~I
the star formation rate is relatively high in this same age range 
(e.g.  Dolphin 2002, fig.~15). In Fornax, too, there 
is significant star formation in this age range 
(e.g. Coleman \& de Jong 2000, fig.~7).

The  case of Sculptor is particularly interesting in that it is frequently said to 
have only a very old population. Nevertheless, the presence of a 554-day
carbon-rich Mira  indicates a 
significant population of age $\sim1-2$ Gyr.  
Interestingly, the model of Revaz et al. 
(2009, fig.~13) does in fact show a small population in the 0-2 Gyr age range. 
Earlier work (Tolstoy et al. 2001, fig.~14) had this population
at $\sim 4$ Gyr.

In both Leo~I and Fornax there has been significant star formation over the whole period,
1 -- 12 Gyr. Thus, the range of Mira periods found in these galaxies is
expected if period depends on age. This does not itself establish an age--period relation.
However, as was previously pointed out (Menzies et al. 2010) two of the Leo~I Miras fall in the outer parts
of the galaxy where, as shown by Mateo et al. (2008), stars on the extended AGB are rare 
and where the main stellar population is older than in the inner parts. 
These two stars have periods of 191 and 283 days  and 
are among the shorter period stars as we would expect
for an older population. 

Sculptor is particularly interesting in this regard. Although,
as we have just seen, the presence of a 554-day Mira is consistent with the small $\sim1$ Gyr
population of the model, there is no evidence of an intermediate population between that and 
the main population at $\sim10$ Gyr and older. Since the evidence points against a
carbon-rich Mira 
of 189 day period belonging to a young population we must assign it to this old population.

Carbon stars in the Galactic halo are considered to be either intermediate age stars
belonging to infalling streams or else old, low mass objects in binaries that have
acquired their carbon from an initially more massive companion. This latter mechanism
is reasonable since only a relatively low percentage of halo giants are carbon rich.
This does not seem to be a viable explanation for the short period carbon-rich Miras in the
dwarf spheroidals unless there are undetected short period oxygen-rich Miras in them, which is very unlikely.
%Evolutionary models of low mass stars do not usually produce carbon stars. 
Evolutionary models of low mass stars do not usually produce carbon
stars (e.g. Iben \& Renzini 1983).
The exact details of the mass and metallicity at which carbon stars do
form depends on the models and we note that recent work (e.g. Karakas 2010; Suda \& Fujimoto 2010)
suggests that low mass stars with low metallicity may become carbon rich.

%However,
%this may be possible in stars of low metallicity (see Karakas 2010; Suda \& Fujimoto 2010).

%POSSIBLY ADD REF TO SUDA/FUJIMOTO (ARXIV: 1002.0863?\\
 It is also of
interest to note that whereas the intermediate age 554-day Mira is centrally located in
Sculptor, the old 189 day one is outside the main concentration of stars in the galaxy.
%WE ALSO NEED TO POINT OUT WHICH VARIABLE
%IS WHICH IN THE CAPTION TO FIG 3  AND TO PUT THE VARIABLE STAR NAMES IN TABLE 2.\\
%CAN WE SAY THE VARIABLE IN PHOENIX IS CENTRALLY LOCATED OR THE THE MAIN BODY?
A similar Mira in Phoenix, with period 425 days, is only about 1.3 arcmin from the centre of that galaxy, compared with the overall extent of about 8.7 arcmin (van de Rydt et al. 1991). 

The relative numbers of Miras in the various galaxies  is roughly what we might expect
from the total luminosities of the galaxies and estimates of the age distribution
of their stars. Thus, data in Amorisco \& Evans (2010) suggest the total luminosities
of Fornax and Leo~I are $14\pm 4$ and $3.4\pm 1.1$ in units of $10^{6}$ solar luminosities
and both have had ongoing star formation from early times until quite recently. However,
the intermediate age population is relatively stronger compared to the very old population
in Leo~I than in Fornax (e.g. Dolphin 2002, fig.~15; Revaz 2009, fig.~13). Thus the roughly
equal number of Miras is understandable. Sculptor is less luminous ($\sim 1.4 \pm 0.6$ in the
same units,) while Phoenix has a similar luminosity, and one expects fewer Miras in these galaxies.
Both have a dominant old population and another increase in star formation later
($\lesssim 5$ Gyr in the case of Phoenix, $\lesssim 2$ Gyr in the case of Sculptor
(Young et al. 2007; Revaz et al. 2009), though with the old population being
relatively stronger in the case of Sculptor. Within the constraints of small number
statistics these results for Miras in Local Group galaxies are consistent with a
general period--age relation for Miras and the known properties of the galaxies.

\section{ Carbon-rich Miras: metallicities, mass loss and luminosities}

The extent to which AGB stars produce dust at low metallicities is of importance
among other things for an understanding of dust formation in the early universe.

%Observations in the Magellanic Clouds and in our own Galaxy 
%(e.g. Sloan et al. 2008) have indicated that whilst the
%dust production from oxygen-rich AGB stars decreases with decreasing
%metallicity this is not the case for carbon-rich AGB stars. Furthermore,
%Sloan et al. (2009) have used Spitzer spectra to deduce that
%MAG 29  is producing significant amounts of dust.

There is a general expectation that, other things being equal, the
dust production in oxygen-rich AGB stars will decrease with decreasing
metallicity since the amount of available material will decrease.
In AGB carbon stars, however, this restriction does not necessarily apply
since fresh carbon is transported to the stellar atmosphere by dredge-up.
Tests of these expectations using observations in the Galaxy and the Magellanic
Clouds have not been entirely conclusive (see Sloan et al. 2008; van Loon et al.
2008). Nevertheless, this and our other papers on dwarf spheroidals have shown
the presence of carbon Miras with thick dust shells in metal-poor systems.
Furthermore, Sloan et al. (2009)  have used Spitzer Space Telescope spectra to deduce that
Sculptor MAG29 is producing significant amounts of dust and have suggested
on this basis that evolved low metallicity stars may make a significant
contribution to dust production in the early Universe.

%Observations in the Magellanic Clouds and in our own Galaxy
%(e.g. Sloan 2008) suggest that dust production in AGB stars decreases
%with decreasing metallicity in oxygen-rich objects but not
%in carbon-rich ones.  This is qualitatively consistent with
%expectation since in oxygen-rich stars the available material for
%dust production decreases with decreasing metallicity. In carbon-rich
%stars, on the other hand, new carbon is transported to the stellar
%atmosphere by dredge up. Furthermore, Sloan et al. (2009) have used Spitzer Space Telescope
%spectra to deduce that Sculptor MAG29 is producing significant amounts
%of dust and have suggested  on this basis that evolved low
%metallicity stars may make a significant contribution to dust
%production in the early Universe.

Our present discovery that this star is a long period Mira allows us to further discuss
this matter together with our work on Miras in other dwarf spheroidals. On the 
basis of the discussion of the last section, MAG29, with a period of 554 days, will be 
a member of the younger population in Sculptor. This has a mean metallicity of 
[Fe/H] $\sim-1.4$ (Tolstoy et al.). This star is similar to the 
longest period Mira in Leo~I, with a period of 523 days, and is thus of similar age.  
Gullieuszik et al. (2009) find from Ca II triplet observations of red giants that Leo~I 
has a rather small dispersion in metal abundance, with [M/H]= --1.2, which they take 
as equivalent to [Fe/H] = --1.4 and thus similar to that inferred for MAG29.
We have already
shown (Menzies at al. 2010) that this star has a significant dust shell since, whilst it 
fits an $M_{bol}$--log P relation, it is too faint, due to shell absorption, for a
$K$--log P 
relation. The difference between the
apparent moduli for the $K$ and $M_{bol}$ relations gives an indication of the optical depth 
of the shell. For the Leo~I star it is 0.8 mag whereas for MAG29 it is 0.4 mag indicating a 
thicker shell in the case of the Leo~I Mira. 
There are other Miras in both Leo~I and
Fornax with similar thick shells indicating high mass-loss. Evidently MAG29 is not particularly
exceptional in this regard.

It is worth noting that in comparing 
mass-loss rates in different stellar systems it is important to consider similar 
types of objects. At least in the case of oxygen-rich Miras the mean mass-loss rate is a function of 
both luminosity (i.e. period) and pulsation amplitude (Whitelock et al. 1987).

Groenewegen, Lan\c{c}on \& Marescaux (2009) have drawn attention to the great strength of
the 1.53 $\mu$m band of C$_2$H$_2$ (acetylene) band in MAG29. Sloan et al. have discussed 
the bands  of this molecule at 7.5 and 13.7 $\mu$m in MAG29 and connect their strength 
with the low metallicity. The 1.53 $\mu$m band is known from the work of  Joyce (1998)
to vary strongly with pulsation phase in the 421-day carbon-rich Mira V~Cyg and comparing his results 
with the AAVSO light curve shows it to be very strong near minimum light. Our infrared 
light curves show that MAG29 was near maximum light when the data of both Groenewegen et
al. 
and Sloan et al. were obtained implying unusual strength
compared with stars so far
observed in this region. However, we might expect it to be of similar strength in
long-period,  dust-enshrouded Miras in other Local Group systems.
It is thus clear that carbon-rich Miras can produce significant amounts
of dust in metal-poor systems. However, it is important to
realize that the carbon-rich Miras in the dwarf spheroidals we have discussed are,
judging from their periods ($\lesssim 500$ days), too old
($\sim$ 1 Gyr) and of too low initial mass ($\sim 2 M_{\odot}$) to be
directly relevant to
the early Universe. At the metallicities
of our Galaxy or the LMC,  Miras with intermediate initial masses and ages
of
relevance for the early Universe
(periods $\gtrsim 1000$ days) are not carbon rich. This is attributed to Hot
Bottom Burning
which turns the carbon into nitrogen. Nevertheless, theory (e.g. Karakas
2010)
suggests that at much
lower metallicities, intermediate mass AGB stars do have $ \rm C/O > 1$
and will thus be carbon stars.

%In our previous papers, on Leo~I, Phoenix and Fornax, we have found no evidence for a significant 
%metallicity effect on the zero-point of either the PL($K$) or the
%PL($M_{bol}$) relations. In the 
%case of Sculptor a mean of the five estimates of the modulus in Table 2 is 19.41.
%The deviations of MAG29 and V544 from this are $+0.16$ and $-0.22$ and are comparable
%with the expected scatter (observational and intrinsic).

%the long period (554-day) variable fits the PL(Mbol) relation  within the 
%expected uncertainties ($\sim0.1-0.2$ mag). The 189-day variable is somewhat discrepant as noted above. 
%As argued above this star belongs to the older population in Sculptor and is likely to be 
%more metal deficient than MAG29, possibly having [Fe/H] as low as $\sim-1.8$ (Tolstoy 2001, 
%Clementini 2005) and this therefore leaves open the possibility of a metallicity effect. 
%However if such exists it would have to affect both the PL(K) and the PL(Mbol) relations by 
%the same amount.

\section{Conclusions}

Infrared monitoring of Sculptor over a four year period has led to
the following results:\\
(1) A comparison with spectroscopically measured abundances shows that
the width of the giant branch is primarily due to a metallicity spread
with extreme metal poor stars ([Fe/H] $<$ --3.4) on the extreme blue
side of the branch.\\
(2) Apart from the two variables discovered, the known carbon stars in
the galaxy
all lie below the TRGB. This contrasts with the situation in other
dwarf spheroidal galaxies. However, an isochrone with [Fe/H] $ \approx -1.1$
and an age of  2 Gyr (Marigo et al. 2008) has an abundance ratio C/O $>$ 1
extending down to the luminosities of these stars.\\
(3) Two carbon-rich Miras are present in Sculptor and give a mean distance modulus of
19.38 mag based on an LMC  modulus of 18.39 mag.\\
(4) A general discussion of the relation of Mira periods to age, together
with other evidence leads to the conclusion that the 554-day period Mira in
Sculptor belongs to the small population with an age of $\sim$ 2 Gyr whilst
the 189-day Mira belongs to the dominant population of age $\sim$ 10 Gyr.\\
(5) The 554-day Mira is identical with the star MAG29 which has been previously
discussed from Spitzer Space Telescope spectra as a low metallicity star with heavy mass loss.
Using our previous work we find that there are several carbon-rich Miras in other
metal-poor dwarf spheroidals with similarly thick circumstellar shells.
Whilst this appears to strengthen the suggestion of Sloan et al. (2009)
that low-metallicity carbon stars could make a significant contribution to
dust formation in the early Universe, this depends on whether young, high-mass
AGB stars can become carbon-rich.

\section{Acknowledgements}

This research has made use of Aladin. 
This publication makes use of data products from the Two Micron All Sky
Survey, which is a joint project of the University of Massachusetts and the
Infrared Processing and Analysis Center/California Institute of Technology,
funded by the National Aeronautics and Space Administration and the National
Science Foundation.
This material is based upon work supported financially by the South 
African National Research Foundation.
We are grateful to our colleagues, Toshihiko Tanab\'e, Yoshifusa Ita, Shogo Nishiyama, Enrico Olivier, Ryo Kandori and Jun Hashimoto, who assisted us or obtained some images of Sculptor 
for us; to Gary Da Costa and Giuseppina Battaglia for providing some useful data; and also to Greg Sloan for sending us the date of the Spitzer observation of MAG29.

\end{document}